\documentstyle[onecolumn,amssym]{mn}

\font\japit = cmti10 at 10truept
\input{epsf.sty}

\title
     [Formulae for Growth Factors In Lambda Universes]
{\vglue-3.0truecm
\centerline{\japit Accepted for publication in Monthly Notices}
\vglue 2.5truecm
\noindent
Formulae for Growth Factors In Expanding Universes Containing Matter and a Cosmological Constant
\author[A. J. S. Hamilton]
     {A. J. S. Hamilton \\
	JILA and Dept.\ Astrophysical \& Planetary Sciences,
	Box 440, U. Colorado, Boulder CO 80309, USA; \\
	\ Andrew.Hamilton@Colorado.EDU; http:$/\!/$casa.colorado.edu/$\sim$ajsh/}
}

\newcommand{\dd}{{\rm d}}	

\newcommand{\aeq}{a_{\rm eq}}

\newcommand{\aap}[2]{A\&A, #1, #2}

\newcommand{\aj}[2]{AJ, #1, #2}
\newcommand{\apj}[2]{ApJ, #1, #2}
\newcommand{\apjs}[2]{ApJS, #1, #2}
\newcommand{\araa}[2]{ARAA, #1, #2}
\newcommand{\mn}[2]{MNRAS, #1, #2}
\newcommand{\nat}[2]{Nature, #1, #2}
\newcommand{\pnas}[2]{Proc.\ Nat.\ Acad.\ Sci., #1, #2}

\newcommand{\Omegam}{\Omega_{\rm m}}
\newcommand{\Omegak}{\Omega_{\rm k}}
\newcommand{\Omegal}{\Omega_\Lambda}

\newcommand{\gffig}{
    \begin{figure}
    \begin{center}
    \leavevmode
    \epsfxsize=2.5in	
    \epsfbox{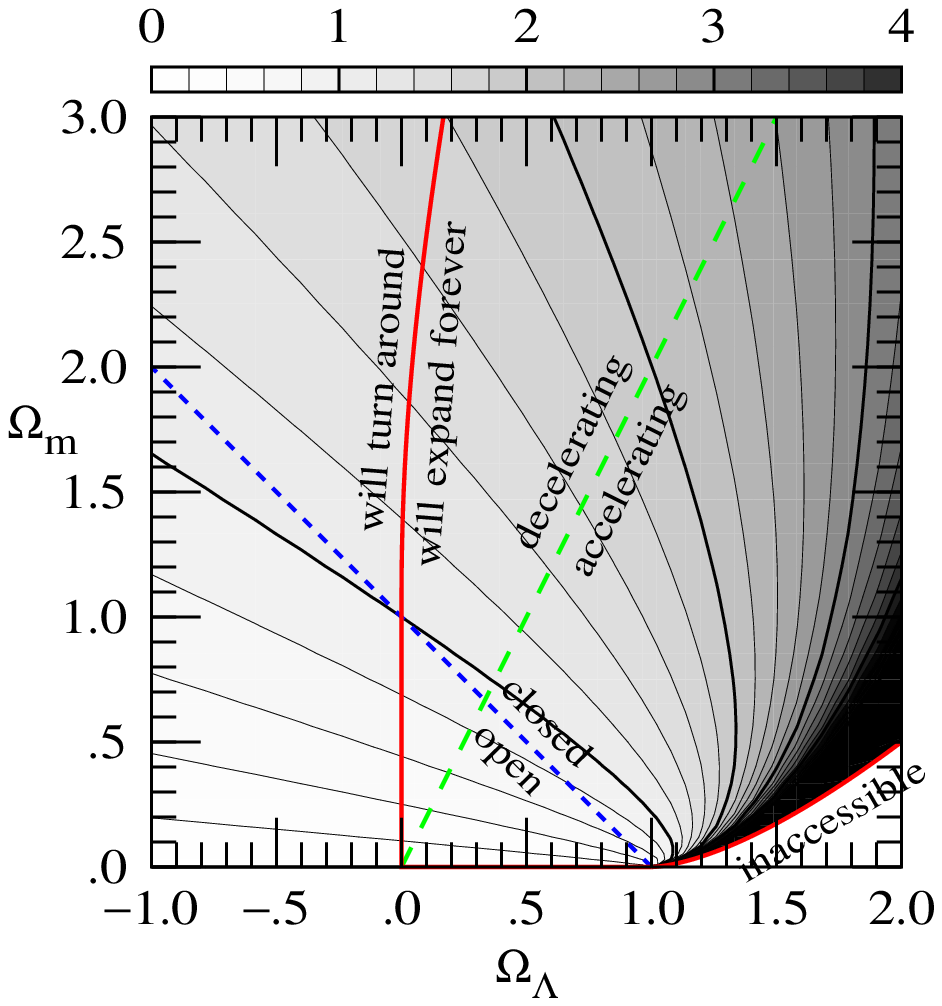}
    \epsfxsize=2.5in	
    \epsfbox{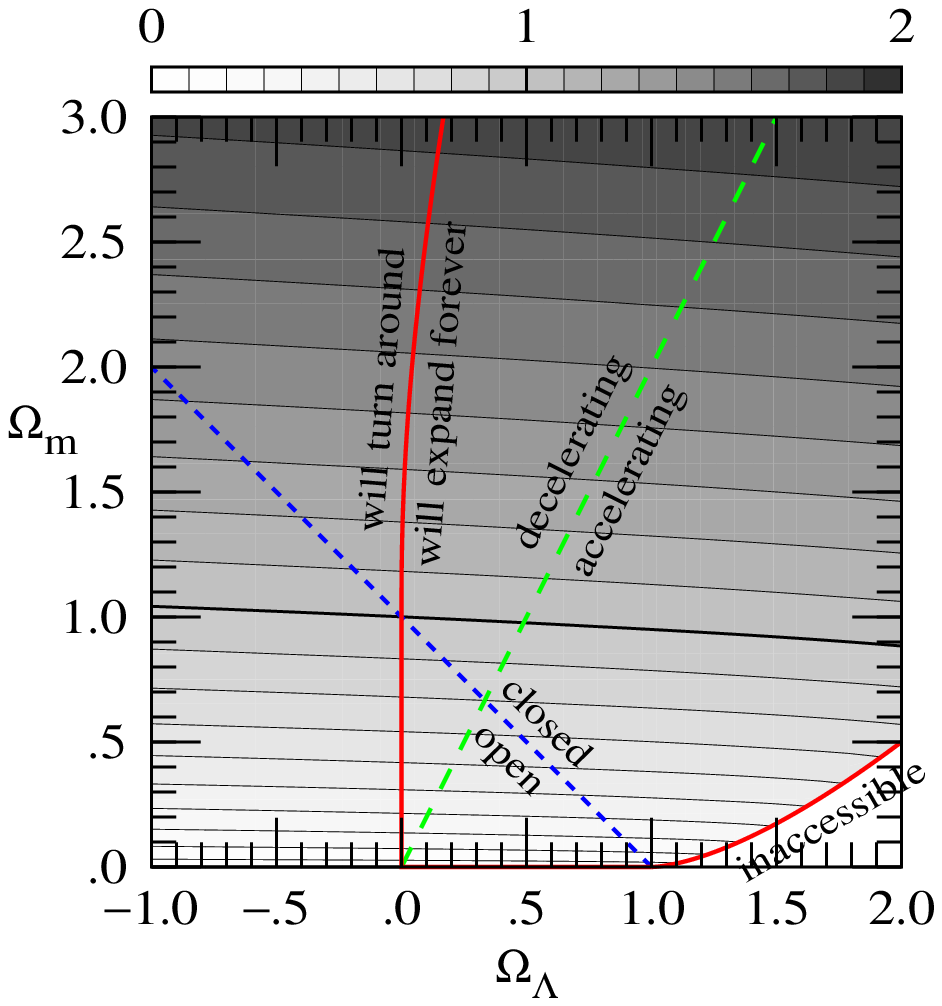}
    \end{center}
    \caption[1]{\small
Contour plots of
(left) the growth factor
$g(\Omegam,\Omegal)$,
and
(right)
the growth rate
$f(\Omegam,\Omegal)$,
in expanding Universes containing matter and cosmological constant
with densities
$\Omegam$ and $\Omegal$
relative to the critical density.
Universes below the $45^\circ$ dashed line
are geometrically open,
while those above are closed.
Universes to the upper left of the long dashed line are decelerating,
while those to lower right are accelerating.
Universes to the left of the almost vertical line near $\Omegal \approx 0$
will eventually turn around,
while Universes to the right will expand forever.
The region at the bottom right corner is physically inaccessible
to Universes that expand from zero
and that contain only matter and cosmological constant.
The boundary between turnaround and eternal expansion,
and the boundary of the inaccessible region,
together form, in the closed case,
the approaching and receding parts of the loiter line.
Starting at $\Omegam = 1$, $\Omegal = 0$,
a Universe can evolve upward and rightward along the approaching part of the
loiter line to Einstein's loiter point
at $\Omegam = 2 \, \Omegal \rightarrow \infty$.
After an indefinite period of hanging around, the Universe
can then either recollapse along the same loiter line,
or else continue into renewed expansion downward and leftward
along the receding part of the loiter line,
the boundary of the inaccessible region.
The growth factor $g$ tends to infinity at the boundary of the
inaccessible region, but only contours up to $g = 5$ are marked.
The contour plot of $g(\Omegam,\Omegal)$ is similar to Figure~7 of
Carroll et al.\ (1992).
    \label{gf}
    }
    \end{figure}
}

\newcommand{\fratfig}{
    \begin{figure}
    \begin{center}
    \leavevmode
    \epsfxsize=2.5in	
    \epsfbox{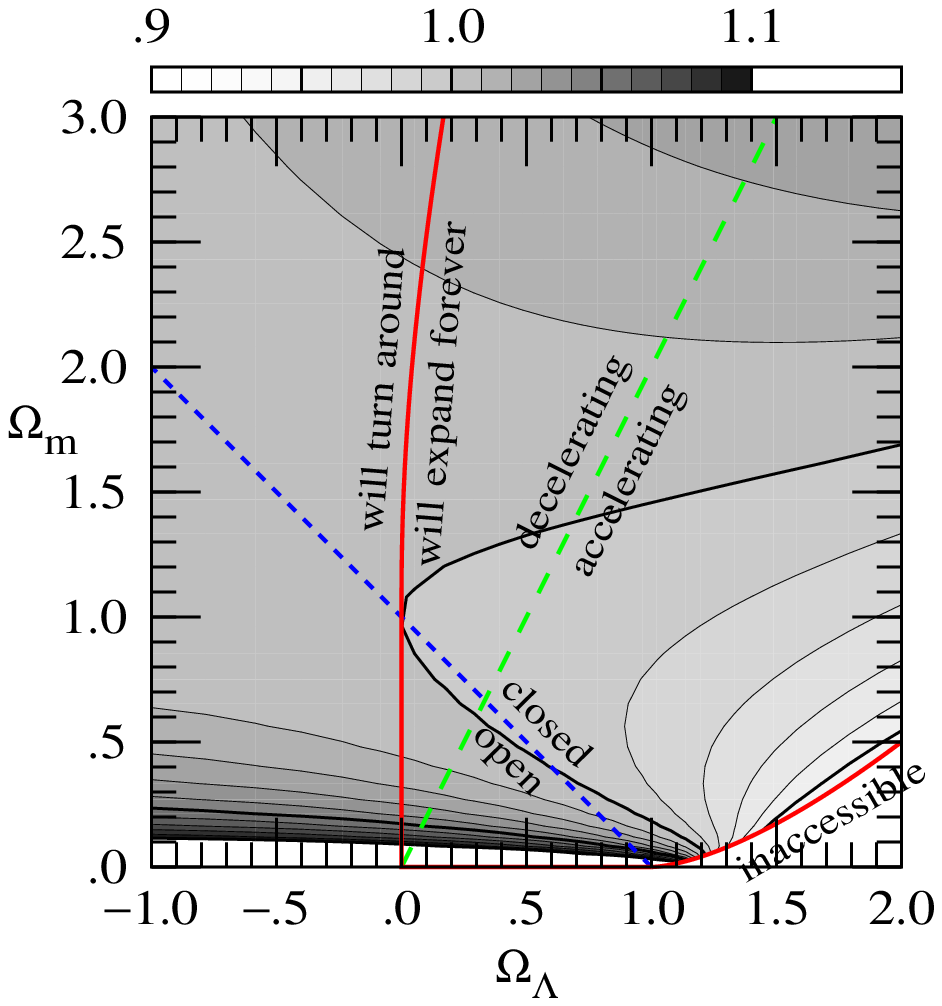}
    \end{center}
    \caption[1]{\small
Contour plot of the ratio
of the Lahav et al.\ (1991) growth rate $f$,
equation~(\ref{fLLPR}),
to the true growth rate.
The plot illustrates that the Lahav et al.\ approximation
works well except for small $\Omegam$.
In particular, the approximation
is accurate to better than 1\% for currently favoured cosmologies,
flat Universes ($45^\circ$ dashed line)
with $\Omegam \ge 0.2$.
For small $\Omegam$ (and any $\Omegal$)
a somewhat better approximation results if
the Lightman \& Schechter (1990) exponent $4/7$
in equation~(\ref{fLLPR}) is replaced with $0.6$,
as in Lahav et al.'s original paper.
The white region in the diagram at $\Omegam \la 0.1$
and $\Omegal < 1$
is where the error in the Lahav et al.\ approximation
(with the $4/7$ exponent) exceeds 10\%;
contours in this region are omitted to avoid confusion.
    \label{frat}
    }
    \end{figure}
}

\begin{document}

\maketitle

\begin{abstract}
Formulae are presented for the linear growth factor $D/a$
and its logarithmic derivative $\dd \ln D/\dd \ln a$
in expanding Friedmann-Robertson-Walker Universes
with arbitrary matter and vacuum densities.
The formulae permit rapid and stable numerical evaluation.
A fortran program is available at
{\tt http://\discretionary{}{}{}casa\discretionary{}{}{}.colorado\discretionary{}{}{}.edu/\discretionary{}{}{}$\sim$ajsh/\discretionary{}{}{}growl/\/}.
\end{abstract}



\section{Introduction}
\label{intro}

The linear growth factor
$g \equiv {D / a}$,
where $D$ is the amplitude of the growing mode
and $a$ is the cosmic scale factor,
determines the normalization of
the amplitude of fluctuations in Large Scale Structure (LSS)
relative to those in the Cosmic Microwave Background (CMB)
(Eisenstein, Hu \& Tegmark 1999 Appendix B.2.2).
Its logarithmic derivative,
the dimensionless linear growth rate
$f \equiv \dd \ln D / \dd \ln a$,
determines the amplitude of
peculiar velocity flows and redshift distortions
(Peebles 1980 \S14;
Willick 2000;
Hamilton 1998).
As such,
the growth factor $g$ and growth rate $f$
are of basic importance in connecting theory and observations
of LSS and the CMB.

In a Friedmann-Robertson-Walker (FRW) Universe containing only
matter and vacuum energy,
with densities $\Omegam$ and $\Omegal$
relative to the critical density,
the linear growth factor is given by\footnote{
Regarded as a function of cosmic scale factor $a$,
the growth factor evolves as
\[
  g(a)
  = {5 \, \Omegam \over 2} {H(a) \over a}
    \int_0^a {\dd a' \over a'^3 H(a')^3}
  = {5 \, \Omegam(a) \, a^2 H(a)^3 \over 2}
    \int_0^a {\dd a' \over a'^3 H(a')^3}
\]
where
$\Omegam(a) = \Omegam a^{-3}/H(a)^2$.
}
(Heath 1977; Peebles 1980 \S10)
\begin{equation}
\label{g}
  g(\Omegam,\Omegal)
  \equiv {D \over a}
  = {5 \, \Omegam \over 2} \int_0^1 {\dd a \over a^3 H(a)^3}
  \ ,
\end{equation}
where $a$ is the cosmic scale factor
normalized to unity at the epoch of interest,
$H(a)$ is the Hubble parameter normalized to unity at $a = 1$,
\begin{equation}
\label{H}
  H(a) =
    \left( \Omegam a^{-3} + \Omegak a^{-2} + \Omegal \right)^{1/2}
  \ ,
\end{equation}
and the curvature density $\Omegak$ is defined to be the density deficit
\begin{equation}
\label{Ok}
  \Omegak \equiv 1 - \Omegam - \Omegal
  \ ,
\end{equation}
which is respectively positive, zero, and negative
in open, flat, and closed Universes.
The normalization factor of $5 \, \Omegam/2$ in equation~(\ref{g})
ensures that $g \rightarrow 1$ as $a \rightarrow 0$.
It follows from equation~(\ref{g}) that
the dimensionless linear growth rate $f$ is related to the growth factor $g$ by
\begin{equation}
\label{f}
  f(\Omegam,\Omegal)
  \equiv {\dd \ln D \over \dd \ln a}
  =
    \mbox{} - 1 - {\Omegam \over 2} + \Omegal
    + {5 \, \Omegam \over 2 \, g}
  \ .
\end{equation}

The fact that the integrand on the right hand side of equation~(\ref{g})
is a rational function of the square root of a cubic
implies that the integral
can be written analytically in terms of elliptic functions.
However, the analytic expressions are complicated,
and have yet to be implemented in any published code.
Explicit analytic expressions
in the special case of a flat Universe, $\Omegak = 0$,
have been given in terms of the incomplete Beta function by
Bildhauer, Buchert \& Kasai (1992)
and in terms of the hypergeometric function by
Matsubara (1995).
Analytic expressions for the luminosity distance
in the general non-flat case are given in terms of elliptic functions by
Kantowski, Kao \& Thomas (2000).

Lahav et al.\ (1991)
give a simple and widely used approximation to the growth rate\footnote{
Lahav et al.\ adopt $\Omega^{0.6}$ rather than $\Omega^{4/7}$,
following Peebles (1980).
The $0.6$ exponent is more accurate for low $\Omegam$,
but $4/7$
(Lightman \& Schechter 1990)
works better elsewhere,
and in particular is more accurate
for currently favoured cosmologies,
flat Universes with $\Omegam \ge 0.2$.
The $4/7$ exponent is therefore currently favoured.
}:
\begin{equation}
\label{fLLPR}
  f(\Omegam,\Omegal)
  \approx
    \Omegam^{4/7} + (1 + \Omegam/2) \, \Omegal/70
  \ .
\end{equation}
From this, together with relation~(\ref{f}),
follows the approximate expression for the growth factor
quoted by Carroll, Press and Turner
(1992):
\begin{equation}
\label{gLLPR}
  g(\Omegam,\Omegal)
  \approx
   {5 \, \Omegam \over
    2 \, \left[ \Omegam^{4/7} - \Omegal
     + (1 + \Omegam/2) (1 + \Omegal/70) \right]}
  \ .
\end{equation}

Given the increasing precision of measurements of fluctuations
in the CMB
(de Bernardis et al.\ 2000;
Hanany et al.\ 2000)
and LSS
(Gunn \& Weinberg 1995; 
York et al.\ 2000; 
Colless 2000)
and the growing evidence favouring a cosmological constant
(Gunn \& Tinsley 1975;
Perlmutter et al.\ 1999;
Riess et al.\ 1998;
Kirshner 1999),
it seems timely to present exact expressions,
suitable for numerical evaluation,
for the growth factor $g$
(hence $f$, through formula~[\ref{f}])
valid for arbitrary values of the cosmological densities
$\Omegam$ in matter and $\Omegal$ in vacuum.

The procedure presented in this paper
is to expand the integral of equation~(\ref{g})
as a convergent series of incomplete Beta functions,
conventionally defined by
\begin{equation}
  B(x,a,b) \equiv \int_0^x t^{a-1} (1-t)^{b-1} \, \dd t
  \ .
\end{equation}
In effect, the method can be regarded as generalizing
Bildhauer et al.'s (1992) formula.
What makes the scheme attractive is that
the Beta functions in successive terms of the series
can be evaluated recursively from each other
through the recursion relations
\begin{equation}
\label{recursea}
  a \, B(x,a,b)
  = x^a (1-x)^b + (a+b) \, B(x,a+1,b)
\end{equation}
\begin{equation}
\label{recurseb}
  b \, B(x,a,b)
  = - x^a (1-x)^b + (a+b) \, B(x,a,b+1)
\end{equation}
\begin{equation}
\label{recurseab}
  a \, B(x,a,b+1)
  = x^a (1-x)^b + b \, B(x,a+1,b)
\end{equation}
the last of which follows from the other two.
In practice, each evaluation of the growth factor $g$
involves from one to seven calls to an incomplete Beta function,
followed by elementary recursive operations.

The resulting numerical algorithm is fast
(provided that $\Omegam$ and $|\Omegal|$ are not huge
-- see \S\ref{collapse})
and stable,
and has been implemented in a fortran package
{\tt grow$\lambda$} available at
{\tt http://\discretionary{}{}{}casa\discretionary{}{}{}.colorado\discretionary{}{}{}.edu/\discretionary{}{}{}$\sim$ajsh/\discretionary{}{}{}growl/\/}.
The {\tt grow$\lambda$} package includes an updated version of
an incomplete Beta function originally written by the author a decade ago.

\section{Formulae}

Figure~\ref{gf}
shows contour plots of the growth factor
$g(\Omegam,\Omegal)$
and growth rate
$f(\Omegam,\Omegal)$
computed from the formulae presented below,
as implemented in the code {\tt grow$\lambda$}.
The results have been checked against numerical integrations
with the Mathematica program.

Perhaps the most striking aspect of these plots,
emphasized by Lahav et al.\ (1991),
is that the growth rate $f$ is sensitive mainly to $\Omegam$,
with only a weak dependence on $\Omegal$.

\gffig

Figure~\ref{frat}
shows the ratio of the Lahav et al.\ (1991) growth rate $f$,
equation~(\ref{fLLPR}),
to the true growth rate.
The Figure illustrates that the Lahav et al.\ approximation
works well except at small $\Omegam$.
The approximation
(with the $4/7$ exponent advocated by Lightman \& Schechter 1990)
works particularly well for currently favoured cosmologies,
being accurate to better than $1\%$ for
flat Universes with $\Omegam = 0.2$--$3.9$.

\fratfig

\subsection{Limiting cases}

The requirement that the Hubble parameter $H(a)$,
equation~(\ref{H}),
be the square root of a positive quantity
for all $a$ from 0 (Big Bang) to 1 (now)
imposes two requirements.
The first is that the matter density be positive
\begin{equation}
\label{conda}
  \Omegam \ge 0
  \ ,
\end{equation}
and the second can be interpreted as a condition that the Universe
not be too closed
\begin{equation}
\label{condb}
  \Omegak \ge \min
  \left[ - {3 \, \Omegam \over 2}
  ,
  - \left( {27 \, \Omegam^2 \Omegal \over 4} \right)^{1/3}
  \right]
  \ .
\end{equation}
If either of the two conditions~(\ref{conda}) or (\ref{condb}) were violated,
then the Hubble parameter $H(a)$ would
go to zero at some finite cosmic scale factor $a$,
indicating that the Universe did not expand from $a = 0$,
but turned around from a collapsing to an expanding phase at some finite $a$.

The limiting values of the growth factor $g$ and growth rate $f$
as the matter density goes zero, $\Omegam \rightarrow 0$, are
\begin{equation}
  g(\Omegam=0,\Omegal) = 0
  \ , \quad
  f(\Omegam=0,\Omegal) = 0
\end{equation}
provided also that $\Omegal < 1$,
in accordance with the condition $\Omegak > 0$ from equation~(\ref{condb}).
Physically, structure cannot grow in a Universe containing only vacuum.

The second condition, equation~(\ref{condb}),
is saturated when
$\Omegak \rightarrow - ( 27 \, \Omegam^2 \Omegal / 4 )^{1/3}$
with
$\Omegak < - 3 \, \Omegam / 2$.
This marks the boundary of the inaccessible region to the bottom right
of Figure~\ref{gf}.
The limiting values of the growth factor $g$ and growth rate $f$
in this case are
\begin{equation}
  g(\Omegam,\Omegal) \rightarrow \infty
  \ , \quad
  f(\Omegam,\Omegal) = \mbox{} - 1 - {\Omegam \over 2} + \Omegal
  \ .
\end{equation}
Physically, the growth factor tends to infinity because
such Universes are in renewed expansion after having spent an indefinite period
of time at Einstein's unstable loitering point,
at $\Omegam = 2 \, \Omegal \rightarrow \infty$.

\subsection{Case $|\Omegak/(1-\Omegak)| \le 1$}
\label{casek}

For small curvature density $\Omegak$,
expand the integral in equation~(\ref{g}) as a power series in $\Omegak$:
\begin{equation}
\label{gk}
  g(\Omegam,\Omegal)
    = {5 \, \Omegam \over 2}
    \int_0^1
    {\dd a \over a^3 (\Omegam a^{-3} + \Omegal)^{3/2}}
    \sum_{n=0}^\infty {(-)^n (3/2)_n \over n!}
    \left( {\Omegak a^{-2} \over \Omegam a^{-3} + \Omegal} \right)^n
\end{equation}
where $(x)_n \equiv x(x+1) ... (x+n-1)$ is a Pochhammer symbol.

Each term of the sum on the right hand side of equation~(\ref{gk})
integrates to an incomplete Beta function.
The cases of positive and negative cosmological constant must be
distinguished.
For positive cosmological constant, $\Omegal > 0$,
\begin{equation}
\label{gkp}
  g(\Omegam,\Omegal)
    = {5 \, \Omegam^{1/3} \over 6 \, \Omegal^{5/6}}
    \sum_{n=0}^\infty {(-)^n (3/2)_n \over n!}
    \left( {\Omegak \over \Omegam^{2/3} \Omegal^{1/3}} \right)^n
    B \left( {\Omegal \over \Omegam+\Omegal},
      {5 \over 6} + {n \over 3}, {2 \over 3} + {2n \over 3} \right)
  \ ,
\end{equation}
while for negative cosmological constant, $\Omegal < 0$,
\begin{equation}
\label{gkm}
  g(\Omegam,\Omegal)
    = {5 \, \Omegam^{1/3} \over 6 \, |\Omegal|^{5/6}}
    \sum_{n=0}^\infty {(-)^n (3/2)_n \over n!}
    \left( {\Omegak \over \Omegam^{2/3} |\Omegal|^{1/3}} \right)^n
    B \left( {|\Omegal| \over \Omegam},
      {5 \over 6} + {n \over 3}, - {1 \over 2} - n \right)
  \ .
\end{equation}
In each of formulae~(\ref{gkp}) and (\ref{gkm}),
incomplete Beta functions must be evaluated for three terms,
and then the remaining Beta functions can be evaluated recursively
from these three,
through the recursion relations (\ref{recursea})--(\ref{recurseab}).
In equation~(\ref{gkp}) ($\Omegal > 0$),
the recursion is stable
from $n=0$ upward, or from large $n$ downward,
according to whether the argument
$\Omegal /(\Omegam + \Omegal)$
is greater than or less than\footnote{
A somewhat lengthy calculation, confirmed by numerical experiment,
shows that the asymptotic (i.e. after many iterations)
point of neutral stability of the recurrence
$B(x,a,b) \rightarrow B(x,a+m,a+n)$,
where $m$ and $n$ are positive or negative integers,
occurs at that unique point $x \in [0,1]$ satisfying
\[
  x^m (1-x)^n = \pm {m^m n^n \over (m+n)^{m+n}}
  \ .
\]
}
$1/3$.
In equation~(\ref{gkm}) ($\Omegal < 0$),
the recursion is stable from $n=0$ upward or large $n$ downward
as the argument
$|\Omegal| / \Omegam$
is greater or less than
$1 + (3/2)[(\sqrt{2}-1)^{1/3} - (\sqrt{2}-1)^{-1/3}] = 0.106$.

How fast do the expansions~(\ref{gkp}) and (\ref{gkm}) converge?
Convergence is determined essentially by the convergence
of the parent expression~(\ref{gk}) at the place where
the expansion variable $\Omegak a^{-2}/(\Omegam a^{-3} + \Omegal)$
attains its largest absolute magnitude over the integration range $a \in [0,1]$.
For $\Omegal > 0$,
the expansion variable attains its largest magnitude at
$a = \min \{ 1 ,  [ \Omegam / ( 2 \, \Omegal ) ]^{1/3} \}$,
while for $\Omegal < 0$,
the expansion variable is always largest at $a = 1$.
Physically, the extremum at $\Omegam a^{-3} = 2 \, \Omegal$ occurs where the
Universe transitions from decelerating to accelerating.
It follows that if $\Omegam \ge 2 \, \Omegal$ (decelerating),
then successive terms of the expansions~(\ref{gkp}) and (\ref{gkm}) decrease by
$\approx \Omegak /(\Omegam+\Omegal)$,
whereas if $\Omegam \le 2 \, \Omegal$ (accelerating),
then successive terms decrease by
$\approx 2^{2/3} \Omegak / ( 3 \, \Omegam^{2/3} \Omegal^{1/3} )$.
Thus $n$ terms of the expansions~(\ref{gkp}) and (\ref{gkm})
will yield a precision of
$\approx [\Omegak /(\Omegam+\Omegal)]^n$
if $\Omegam \ge 2 \, \Omegal$,
or a precision of
$\approx [ 2^{2/3} \Omegak / ( 3 \, \Omegam^{2/3} \Omegal^{1/3} ) ]^n$
if $\Omegam \le 2 \, \Omegal$.

\subsection{Case $|\Omegal/(1-\Omegal)| \le 1$}
\label{casel}

For small cosmological constant $\Omegal$,
expand the integral in equation~(\ref{g}) as a power series in $\Omegal$:
\begin{equation}
\label{gl}
  g(\Omegam,\Omegal)
    = {5 \, \Omegam \over 2}
    \int_0^1
    {\dd a \over a^3 (\Omegam a^{-3} + \Omegak a^{-2})^{3/2}}
    \sum_{n=0}^\infty {(-)^n (3/2)_n \over n!}
    \left( {\Omegal \over \Omegam a^{-3} + \Omegak a^{-2}} \right)^n
  \ .
\end{equation}

Again, each term of the sum on the right hand side of equation~(\ref{gl})
integrates to an incomplete Beta function.
The cases of positive and negative curvature density $\Omegak$ must be
distinguished.
For an open Universe, $\Omegak > 0$,
\begin{equation}
\label{glp}
  g(\Omegam,\Omegal)
    = {5 \, \Omegam^2 \over 2 \, \Omegak^{5/2}}
    \sum_{n=0}^\infty {(-)^n (3/2)_n \over n!}
    \left( {\Omegam^2 \Omegal \over \Omegak^3} \right)^n
    B \left( {\Omegak \over \Omegam + \Omegak},
      {5 \over 2} + 3n , - 1 - 2n \right)
  \ ,
\end{equation}
while for a closed Universe, $\Omegak < 0$,
\begin{equation}
\label{glm}
  g(\Omegam,\Omegal)
    = {5 \, \Omegam^2 \over 2 \, |\Omegak|^{5/2}}
    \sum_{n=0}^\infty {(-)^n (3/2)_n \over n!}
    \left( {\Omegam^2 \Omegal \over |\Omegak|^3} \right)^n
    B \left( {|\Omegak| \over \Omegam},
      {5 \over 2} + 3n , - {1 \over 2} - n \right)
  \ .
\end{equation}
Here the Beta functions in successive terms can be computed recursively
from a single Beta function.
In equation~(\ref{glp}) ($\Omegak > 0)$,
the recursion is stable from $n=0$ upward or large $n$ downward
as the argument
$\Omegak /(\Omegam + \Omegak)$
is greater than or less than $3/4$.
In equation~(\ref{glm}) ($\Omegak < 0$),
the recursion is stable from $n=0$ upward or large $n$ downward
as the argument
$|\Omegak| / \Omegam$
is greater than or less than
$(3/2)[(\sqrt{2}+1)^{1/3} - (\sqrt{2}+1)^{-1/3}] = 0.894$.

The convergence of the expansions~(\ref{glp}) and (\ref{glm})
is determined by the convergence of the parent expansion~(\ref{gl})
at the place where
the expansion variable $\Omegal /(\Omegam a^{-3} + \Omegak a^{-2})$
attains its largest magnitude over the integration range $a \in [0,1]$,
which occurs at $a = 1$ for both positive and negative $\Omegak$.
It follows that
successive terms of the expansions~(\ref{glp}) and (\ref{glm}) decrease by
$\approx \Omegal /(\Omegam+\Omegak)$,
and $n$ terms will yield a precision of
$\approx [\Omegal /(\Omegam+\Omegak)]^n$.

\subsection{Case $|\Omegam/(1-\Omegam)| \le 1$}
\label{casem}

For small matter density $\Omegam$,
a power series expansion of the integral in equation~(\ref{g}) is again
possible, but the integral must be split into two parts
to ensure convergence of the integrand
over the full range $a \in [0,1]$ of the integration variable.
Define the cosmic scale factor $\aeq$ at
matter-curvature `equality' to be where
$|\Omegam \aeq^{-3}/(\Omegak \aeq^{-2} + \Omegal)|
= |\Omegak \aeq^{-2}/(\Omegam \aeq^{-3} + \Omegal)|$,
and similarly that at
matter-vacuum `equality' to be where
$|\Omegam \aeq^{-3}/(\Omegak \aeq^{-2} + \Omegal)|
= |\Omegal/(\Omegam \aeq^{-3} + \Omegak \aeq^{-2})|$.
These conditions reduce to
$\Omegam \aeq^{-3} = \Omegak \aeq^{-2}$ if $\Omegak > 0$,
and $\Omegam \aeq^{-3} = \Omegal$ if $\Omegal > 0$.
A good procedure is to use one of the methods of the previous
two subsections, \S\ref{casek} or \S\ref{casel},
to integrate up to the cosmic scale factor
$\aeq$
at matter-curvature or matter-vacuum `equality',
whichever is later (larger $\aeq$)
since the later epoch yields a more convergent $\Omegam$ series,
and then to complete the integral using the small $\Omegam$ expansion:
\begin{equation}
\label{gm}
  g(\Omegam,\Omegal)
    = 
    G(\aeq)
    + {5 \, \Omegam \over 2}
    \int_{\aeq}^1
    {\dd a \over a^3 (\Omegak a^{-2} + \Omegal)^{3/2}}
    \sum_{n=0}^\infty {(-)^n (3/2)_n \over n!}
    \left( {\Omegam a^{-3} \over \Omegak a^{-2} + \Omegal} \right)^n
  \ .
\end{equation}
The first term on the right hand side of equation~(\ref{gm})
is the integral up to $\aeq$,
evaluated by one of the methods of the previous two subsections:
\begin{equation}
\label{G}
  G(\aeq) \equiv
    {5 \, \Omegam \over 2}
    \int_0^{\aeq}
    {\dd a \over a^3 (\Omegam a^{-3} + \Omegak a^{-2} + \Omegal)^{3/2}}
  =
    {\aeq \over H(\aeq)} g \left[ \Omegam(\aeq),\Omegal(\aeq) \right]
\end{equation}
with
$\Omegam(a) = \Omegam a^{-3} / H(a)^2$,
$\Omegak(a) = \Omegak a^{-2} / H(a)^2$,
$\Omegal(a) = \Omegal / H(a)^2$,
and $H(a)$ given by equation~(\ref{H}).

As in the previous two subsections,
each term of the sum in the second term on the right hand side of
equation~(\ref{gm}) integrates to an incomplete Beta function.
The cases of positive and negative curvature and vacuum densities
must be distinguished.
For an open Universe with a positive cosmological constant,
$\Omegak > 0$ and $\Omegal > 0$,
\begin{equation}
\label{gmpp}
  g(\Omegam,\Omegal)
  =
    G(\aeq)
    + {5 \, \Omegam \over 4 \, \Omegak \Omegal^{1/2}}
    \sum_{n=0}^\infty {(-)^n (3/2)_n \over n!}
    \left( {\Omegam \Omegal^{1/2} \over \Omegak^{3/2}} \right)^n
    \left[
    B \left( {\Omegak a^{-2} \over \Omegak a^{-2} + \Omegal},
      1 + {3n \over 2} , {1 \over 2} - {n \over 2} \right)
    \right]_{a=\aeq}^1
  \ ,
\end{equation}
while for an open Universe with a negative cosmological constant,
$\Omegak > 0$ and $\Omegal < 0$,
\begin{equation}
\label{gmpm}
  g(\Omegam,\Omegal)
  =
    G(\aeq)
    + {5 \, \Omegam \over 4 \, \Omegak |\Omegal|^{1/2}}
    \sum_{n=0}^\infty {(-)^n (3/2)_n \over n!}
    \left( {\Omegam |\Omegal|^{1/2} \over \Omegak^{3/2}} \right)^n
    \left[
    B \left( {|\Omegal| \over \Omegak a^{-2}},
      {1 \over 2} - {n \over 2} , - {1 \over 2} - n \right)
    \right]_{a=\aeq}^1
  \ .
\end{equation}
The expression for a closed Universe with positive cosmological constant,
$\Omegak < 0$ and $\Omegal > 0$,
turns out never to be useful, but for reference it is
\begin{equation}
\label{gmmp}
  g(\Omegam,\Omegal)
  =
    G(\aeq)
    + {5 \, \Omegam \over 4 \, |\Omegak| \Omegal^{1/2}}
    \sum_{n=0}^\infty {(-)^n (3/2)_n \over n!}
    \left( {\Omegam \Omegal^{1/2} \over |\Omegak|^{3/2}} \right)^n
    \left[
    B \left( {|\Omegak| a^{-2} \over \Omegal},
      1 + {3n \over 2} , - {1 \over 2} - n \right)
    \right]_{a=\aeq}^1
  \ .
\end{equation}
The fourth option,
a closed Universe with negative cosmological constant,
$\Omegak < 0$ and $\Omegal < 0$,
does not yield a convergent expansion in $\Omegam$.
The small $\Omegam$ expressions~(\ref{gmpp})--(\ref{gmmp})
are more complicated than the small $\Omegak$ and small $\Omegal$ expressions
obtained in the previous two subsections,
since equations~(\ref{gmpp})--(\ref{gmmp})
each involve a sum of not one but three expansions,
one to evaluate the first term on the right hand side,
one to evaluate the second term at $a = \aeq$,
and a third to evaluate the second term at $a = 1$.

It will now be argued that
a small $\Omegam$ expansion is advantageous only in the case where
matter-vacuum equality occurs after matter-curvature equality.
In particular, this has the consequence that
the expansion~(\ref{gmmp}) is never useful.
Suppose that matter-curvature equality,
$|\Omegam \aeq^{-3}/(\Omegak \aeq^{-2} + \Omegal)|
= |\Omegak \aeq^{-2}/(\Omegam \aeq^{-3} + \Omegal)|$,
occurs after matter-vacuum equality.
Then the first term in equation~(\ref{gm}),
$G(\aeq)$,
would be evaluated by the small $\Omegak$ method,
equation~(\ref{gkp}) or (\ref{gkm}).
However,
if matter-curvature equality occurs after matter-vacuum equality,
then it is also true that matter-curvature equality occurs after
(larger cosmic scale factor $a$)
the transition, at $\Omegam a^{-3} = 2 \, \Omegal$,
from a decelerating to an accelerating Universe.
As discussed in the last paragraph of \S\ref{casek},
the convergence of the small $\Omegak$ expansions~(\ref{gkp}) and (\ref{gkm})
are then determined by the convergence of the parent equation~(\ref{gk})
at the deceleration-acceleration transition,
not by its convergence at a later epoch.
Thus,
if matter-curvature equality occurs after matter-vacuum equality,
then there is no point in splitting the integral at $\aeq$
as in equation~(\ref{gm});
one might as well use the small $\Omegak$ expressions~(\ref{gkp}) or (\ref{gkm})
all the way to $a = 1$, since they converge just as fast
at $a = 1$ as at $a = \aeq$.
In fact the small $\Omegak$ expressions~(\ref{gkp}) and (\ref{gkm})
are computationally faster than the small $\Omegam$
expansions~(\ref{gmpp})--(\ref{gmmp}),
because the former involve a single sum whereas the latter involve three.

The conclusion from the previous paragraph is that
a small $\Omegam$ expansion is advantageous only in the case where
matter-vacuum equality occurs after matter-curvature equality.
Examination of evolution in the $\Omegam$--$\Omegal$ plane reveals that 
matter-vacuum equality happens after matter-curvature equality
only if $\Omegak > 0$
(conversely,
matter-curvature equality happens after matter-vacuum equality
only if $\Omegal > 0$).
Thus, in the cases where the small $\Omegam$ expansion is useful,
the relevant expressions to use are~(\ref{gmpp}) or (\ref{gmpm})
with the first term,
$G(\aeq)$,
being evaluated by the small $\Omegal$ expansion~(\ref{glp}) with $\Omegak > 0$.

The Beta functions in successive terms of the expansions in the second
term on the right hand sides of equations~(\ref{gmpp})--(\ref{gmmp})
can be computed recursively from two Beta functions.
In equation~(\ref{gmpp}) ($\Omegak > 0$ and $\Omegal > 0$),
the recursion is stable from $n=0$ upward or large $n$ downward
as the argument
$\Omegak a^{-2} /(\Omegak a^{-2} + \Omegal)$
(with $a = \aeq$ or $1$)
is greater than or less than
$(3/2)[(\sqrt{2}+1)^{1/3} - (\sqrt{2}+1)^{-1/3}] = 0.894$,
the same as for equation~(\ref{glm}).
In equation~(\ref{gmpm}) ($\Omegak > 0$ and $\Omegal < 0$),
the recursion is stable from $n=0$ upward
in all cases.
In equation~(\ref{gmmp}) ($\Omegak < 0$ and $\Omegal > 0$),
the recursion is stable from $n=0$ upward or large $n$ downward
as the argument
$|\Omegak| a^{-2} / \Omegal$
is greater than or less than $3/4$,
the same as for equation~(\ref{glp}).

The convergence of the expansions~(\ref{gmpp})--(\ref{gmmp})
is determined by the convergence of the parent expansion~(\ref{gm})
at the place where
the expansion variable $\Omegam a^{-3} /(\Omegak a^{-2} + \Omegal)$
attains its largest magnitude over the integration range $a \in [\aeq,1]$,
which occurs at $a = \aeq$ in all cases.
It follows that
successive terms in the expansions~(\ref{gmpp})--(\ref{gmmp}) decrease by
$\approx \Omegam \aeq^{-3} /(\Omegak \aeq^{-2} + \Omegal)$,
and $n$ terms will yield a precision of
$\approx [\Omegam \aeq^{-3} /(\Omegak \aeq^{-2} + \Omegal)]^n$.

\subsection{Which formula to use?}

The various formulae~(\ref{gkp}), (\ref{gkm}),
(\ref{glp}), (\ref{glm}), and (\ref{gmpp})--(\ref{gmmp})
converge over overlapping ranges of $\Omegam$ and $\Omegal$.
A sensible strategy would be
to choose the expression that converges most rapidly.

If $\Omegak \le 0$ (closed Universe),
then use the
small $\Omegak$ (\S\ref{casek})
or
small $\Omegal$ (\S\ref{casel})
methods
as
$|\Omegak/(1-\Omegak)|$
or
$|\Omegal/(1-\Omegal)|$
is smallest.

If $\Omegak > 0$ (open Universe),
then use the
small $\Omegak$ (\S\ref{casek}),
small $\Omegal$ (\S\ref{casel}),
or small $\Omegam$ (\S\ref{casem})
methods as
$|\Omegak/(1-\Omegak)|$,
$|\Omegal/(1-\Omegal)|$,
or
$|\Omegam/(1-\Omegam)|$
is smallest.
If $|\Omegam/(1-\Omegam)|$ is smallest,
then determine which occurs later (larger $\aeq$),
matter-curvature equality
$|\Omegam \aeq^{-3}/(\Omegak \aeq^{-2} + \Omegal)|
= |\Omegak \aeq^{-2}/(\Omegam \aeq^{-3} + \Omegal)|$,
or
matter-vacuum equality
$|\Omegam \aeq^{-3}/(\Omegak \aeq^{-2} + \Omegal)|
= |\Omegal/(\Omegam \aeq^{-3} + \Omegak \aeq^{-2})|$.
If the former, then revert to using the small $\Omegak$ method, \S\ref{casek}.
If the latter, then use one of the small $\Omegam$
expressions~(\ref{gmpp}) or (\ref{gmpm})
with the first term,
$G(\aeq)$, equation~(\ref{G}),
evaluated by the small $\Omegal$ expression~(\ref{glp}).


\section{Collapsing Universe}
\label{collapse}

The procedure proposed in this paper fails for Universes that
are collapsing rather than expanding.
As a Universe approaches turnaround,
the Hubble parameter, and consequently the critical density, approaches zero,
causing one or more of $\Omegam$, $\Omegak$, and $\Omegal$
to tend to $\pm \infty$.
In such cases the expansions presented herein converge ever more slowly.

It is not clear how to adapt the method to deal with Universes
near turnaround and thereafter, or indeed whether this is possible.

%

\section{Summary}

Formulae suitable for numerical evaluation have been presented
for the linear growth factor $g(\Omegam,\Omegal) \equiv D/a$
and its logarithmic derivative $f(\Omegam,\Omegal) \equiv \dd \ln D/\dd \ln a$
in expanding FRW Universes
with arbitrary matter and vacuum densities $\Omegam$ and $\Omegal$.
A fortran package {\tt grow$\lambda$} implementing these formulae
is available at
{\tt http://\discretionary{}{}{}casa\discretionary{}{}{}.colorado\discretionary{}{}{}.edu/\discretionary{}{}{}$\sim$ajsh/\discretionary{}{}{}growl/\/}.

\section*{Acknowledgements}

The author thanks Daniel Eisenstein for helpful information and references,
Gil Holder and Ioav Waga for
pointing out an erroneous, now eliminated, section in the original manuscript,
Matthew Graham for finding an important bug,
and Paul Schechter for urging the merits of $4/7$.
This work was supported by
NASA ATP grant NAG5-7128.

\end{document}